\documentclass{article}
\usepackage{graphicx,amssymb, amsmath}
\title{       Charge density wave in the spin ladder of 
              Sr$_{14-x}$Ca$_x$Cu$_{24}$O$_{41}$ }

\author{Krzysztof Wohlfeld,$^{a}$ Andrzej M. Ole\'{s},$^{a,b}$ George A. Sawatzky $^{c}$ 
\\
\scriptsize{$^a$Marian Smoluchowski Institute of Physics, Jagellonian University} \\
\scriptsize{Reymonta 4, PL-30059 Krak\'ow, Poland }\\
\scriptsize{$^b$Max-Planck-Institut f\"ur Festk\"orperforschung}\\
\scriptsize{Heisenbergstrasse 1, D-70569 Stuttgart, Germany}, \\
\scriptsize{$^c$Department of Physics and Astronomy, University of British Columbia} \\
\scriptsize{Vancouver B. C. V6T-1Z1, Canada}}
\begin{document}
\date{}
\maketitle
\begin{abstract}
We consider a multiband charge transfer model for a single spin 
ladder describing the holes in Sr$_{14-x}$Ca$_x$Cu$_{24}$O$_{41}$. 
Using Hartree-Fock approximation we show how the charge density wave, 
with its periodicity dependent on doping as recently observed in the 
experiment, can be stabilized by {\it purely\/} electronic many-body 
interactions.
\end{abstract}
A prerequisite for the understanding of high $T_c$ superconducting 
cuprates (HTSC) is to describe the normal phase and its possible 
instabilities which may lead to other ordered phases that compete with 
the superconducting one. As recently investigated experimentally 
\cite{Rus05}, such a competing phase in the hole doped weakly coupled 
spin ladder system Sr$_{14-x}$Ca$_x$Cu$_{24}$O$_{41}$ (SCCO) is the 
electronic charge density wave (CDW). It is well known that CDW could 
in general be stabilized due to either: 
  ($i$) strong long-range or at least intersite Coulomb interaction, or 
 ($ii$) the Peierls instability, or 
($iii$) charge stripe formation in the single-band ($t$-$J$) model
        \cite{Dag92}. 
However, the long-range Coulomb interaction is largely suppressed in 
HTSC \cite{Gra92} and the Peierls origin of the CDW in SCCO is excluded 
\cite{Rus05}.
Furthermore, the Zhang-Rice \cite{Zha88} derivation of the single band 
$t$-$J$ Hamiltonian for HTSC is still controversial \cite{Mac05} and 
makes it questionable to explain charge order (CO) in SCCO within the 
$t$-$J$ model.
Hence it is puzzling to investigate whether: 
 ($i$) the $d-p$ multiband charge transfer model can explain the onset 
     of the CO in this compound by on-site Coulomb interactions 
     {\it alone}, and 
($ii$) the calculated CDW periodicity agrees with the one observed 
     experimentally and hence could give strong theoretical support for 
     the CDW in SCCO as induced by electronic interactions.

\begin{figure}[t!]
\includegraphics[width=7.5cm]{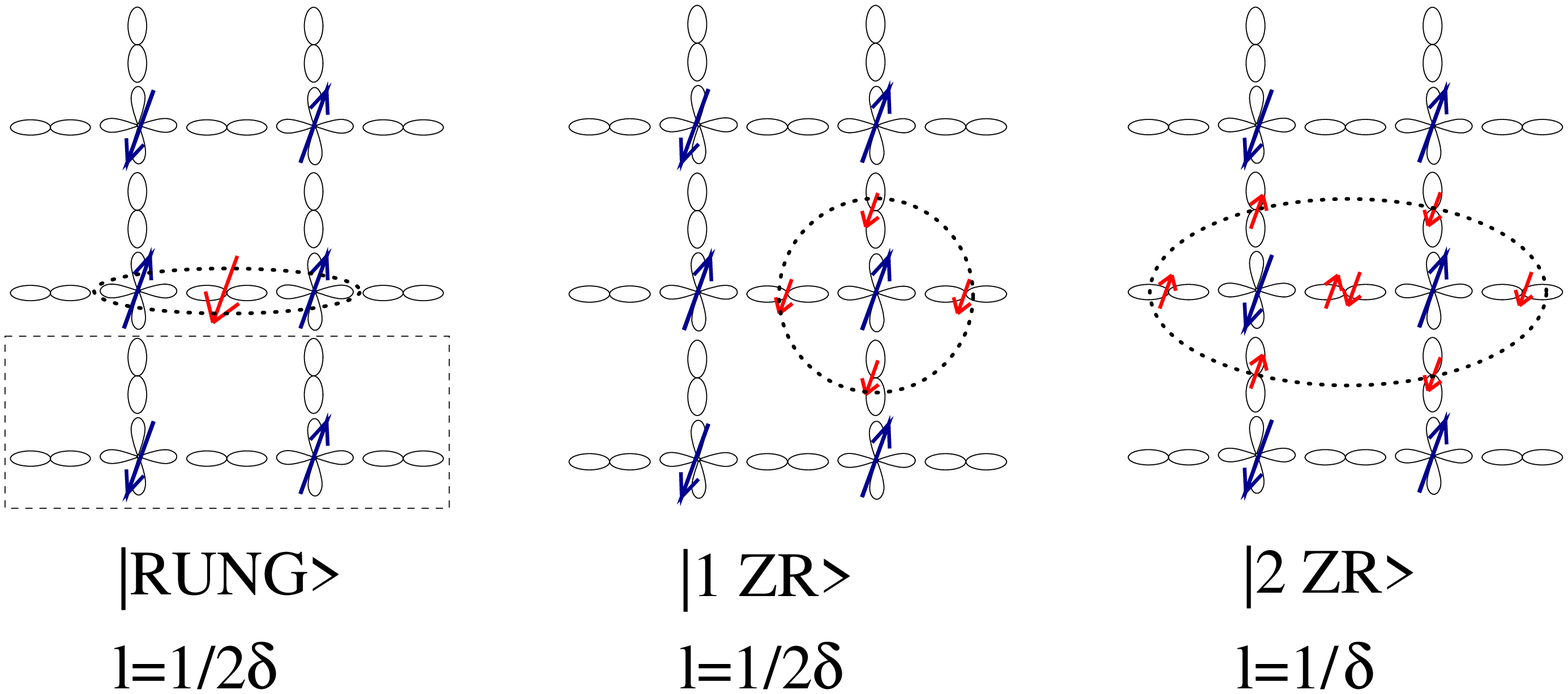}
\caption[]{\small{
Three different CO states in the AF background of a single ladder. 
Large (small) arrows represent the hole spins for +1.0 (+0.25) charge. 
The red arrows depict spins of doped holes. The dotted line shows the 
bound state. For the $|$RUNG$\rangle$ state the Cu$_2$O$_5$ unit cell 
is shown by rectangle.} 
}
\label{fig:fig1} 
\end{figure}

The theoretical model for a decoupled single ladder [the weak coupling 
between the ladders can be neglected in leading order] includes seven 
orbitals per Cu$_2$O$_5$ ladder unit cell (within Cu$_2$O$_3$ plane), 
and is similar to that developed earlier for CuO$_2$ planes of HTSC 
\cite{Ole87}. The unit cell consists of: 
two Cu($3d_{x^2-y^2}\equiv d$) orbitals on two legs of the ladder, 
one bridge O($2p_\sigma\equiv b$) orbital in the inner part of 
each ladder rung, 
two O($2p_\sigma\equiv p$) orbitals on the ladder legs, and two other 
ones on outer parts of the rung (also labelled as $p$); 
cf. Fig. \ref{fig:fig1}. The $d$ and $p$ orbitals belong to two sets 
for the right ($R$) and left ($L$) leg, respectively. 
We include large on-site Coulomb repulsion on copper ($U$) and oxygen 
($U_p$), the charge transfer energy for the $p$ orbitals, 
$\Delta=\varepsilon_{p}-\varepsilon_d$, and for the $b$ orbitals, 
$\varepsilon=\varepsilon_{b}-\varepsilon_d$. For convenience we set 
$\varepsilon_d=0$ and use the hole notation. Hence we consider 
the Hamiltonian, 
\begin{align}
H=&
\Big(-\!\!\!\!\!\sum_{m,j\in R,L,\sigma} 
     \!\!t_{mj} d^\dag_{m\sigma}p_{j\sigma}\!
-\!\!\!\!\sum_{m\in R,L;i\sigma} 
   \!\!t_{mi} d^\dag_{m\sigma}b_{i\sigma}+\mbox{h.c.}\!\Big) 
+\Delta\sum_{j\in R,L}n_{pj} \nonumber \\
&+ \varepsilon\sum_{i}n_{bi} 
 +  U\sum_{m\in R,L} n_{m\uparrow} n_{m\downarrow}           
+ U_p\Big(\sum_{i}       n_{bi\uparrow}n_{bi\downarrow} 
           +\sum_{j\in R,L}n_{pj\uparrow}n_{pj\downarrow}\Big),  
\label{eq:1}
\end{align}
where $t_{mj}$, $t_{mi}$ are the hopping elements between nearest 
neighbor pairs of $\{p,d\}$ orbitals which include the respective phases, 
and $n_{m}=\sum_{\sigma}n_{m\sigma}$, $n_{pj}=\sum_{\sigma}n_{pj\sigma}$,
$n_{bi}=\sum_{\sigma}n_{bi\sigma}$ are particle number operators.

The model parameters ($U=9$ eV, $U_p=4.2$ eV, $\Delta=3.5$ eV) follow from 
ref. \cite{Gra92}, whereas $\varepsilon$ is somewhat smaller than $\Delta$ 
\cite{Rus06}. Present value of $t=1$ eV is motivated by the experimental 
value of the superexchange constant $J$ \cite{Ecc98}, though our results 
turn out to be valid for large range of the hopping parameter $t$.
SCCO is a self-doped compound with $1<n<1.33$ (number of holes/Cu ion in 
the ladder) \cite{Rus06}. In the limit $t\ll U$ one finds one hole in each 
$d$ orbital and the antiferromagnetic (AF) order. Depending on the ratio 
$\varepsilon/\Delta$ the doped holes go either to a linear combination of 
$p/b$ orbitals, or to the $b$ orbitals alone. Hence, in the regime of 
$t\ll\{U,\Delta\}$ the kinetic exchange could favour the formation of: 
 ($i$) Zhang and Rice (ZR) bound state \cite{Zha88}, or 
($ii$) a {\it rung} bound state with the hole being doped into $b$ orbital 
and having an opposite spin to those of two neighboring holes in $d$ 
orbitals. This means that for a commensurate hole doping $\delta=n-1$ 
these states would naturally form a CDW of period $l$ 
(Fig. \ref{fig:fig1}), unless such bound states were not destabilized by 
increasing bandwidth.
\begin{figure}[t!]
\includegraphics[width=7.5cm]{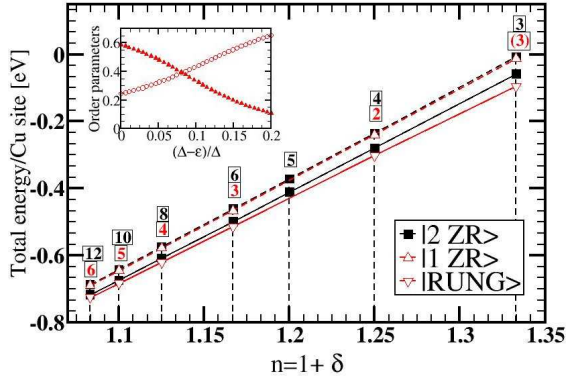}
\caption[]{\small{
Total energy as a function of $n$ for 
$\varepsilon=3.5$ eV (dashed line) and 
$\varepsilon=3.0$ eV (solid line). 
The red (black) color describes the states with one doped hole 
(two holes) per rung. The bottom (top) number in each rectangle 
depicts the CDW period for the lowest (higher) energy state. 
The inset shows the dependence on $\varepsilon$, for $n=1.25$, of the 
order parameters: 
$|$RUNG$\rangle$ (circles; $\langle n_{b,i=2n}\rangle$), and 
$|$1ZR$\rangle$ [triangles; $\langle n_{L}-n_{R}\rangle$, where $n_{L}$ 
($n_{R}$) is the total hole density in the left (right) $p$ orbitals which 
belong to the rung with enhanced hole density].}
} 
\label{fig:fig2}
\end{figure}

In order to verify such a mechanism of CDW formation we studied the 
energies of three different CO states (Fig. \ref{fig:fig1}) determined
self-consistently in the Hartree-Fock approximation, using 
$60\times 7$ or $64\times 7$ clusters. Stable CO {\it insulating} 
states have lower energies than a homogeneuos AF state (not shown).

For $\varepsilon=\Delta$ one finds that:
  ($i$) both of the ZR states are stable whereas $|$RUNG$\rangle$ is 
        unstable, 
 ($ii$) $|$1ZR$\rangle$ state is slightly more favoured 
        energetically than the $|$2ZR$\rangle$ state,
($iii$) the CDW periodicity in the ground state is $l=1/(2\delta)$.
With decreasing ratio $\varepsilon/\Delta$ the $|$1ZR$\rangle$ state 
evolves into a $|$RUNG$\rangle$ state which makes the CDW with one hole 
per rung even more robust. In this case one gains  
$(\Delta-\varepsilon)(\langle n_b \rangle-0.5)$ on-site energy, 
where $\langle n_b\rangle\cong 1$ is the number of holes in $b$ orbital 
in the $|$RUNG$\rangle$ state.

We obtained that for $n=1.2$ the CDW with period $l=5$ is stable, and 
for $n=1.33$ the CDW with less charge on every third rung [$l=(3)$] is 
stable, which both agree with the experiment for $x=0$ [$n=1.2$] and 
$x=11$ [$n=1.3$] in SCCO \cite{Rus06}. Though, for $x=4$ [$n=1.25$] the 
CO has not been observed. We suggest that in this case the CO could be 
destroyed due to weak interladder interaction, as the topology of the 
arrangement of the ladders in the plane does not allow for a 
homogeneous order with an even period. 

In summary, we have shown how the CDW can be stabilized in SCCO in the 
charge transfer model for a Cu$_2$O$_5$ ladder due to on-site Coulomb 
interactions. In the relevant range of parameters the Fermi surface is 
unstable and the CO insulating state is formed by ZR or rung states. 
The obtained periodicities of the CO agree qualitatively with the 
experimental data.

\footnotesize{{\bf Acknowledgments} This work was supported by the Polish Ministry of Science and 
Education Project No. 1~P03B~068~26.}

\end{document}